\documentclass[twocolumn,10pt,prb,aps,showpacs,superscriptaddress,preprintnumbers,amsmath,amssymb]{revtex4-1}

\usepackage{graphicx}
\usepackage{dcolumn}
\usepackage{bm}
\usepackage{eqnarray}
\usepackage{color}
\usepackage{bigstrut}
\usepackage{tabularx}
\usepackage{float}
\usepackage[]{color}
\usepackage{silence}
\newcolumntype{L}[1]{>{\raggedright\arraybackslash}p{#1}}
\newcolumntype{C}[1]{>{\centering\arraybackslash}p{#1}}
\newcolumntype{R}[1]{>{\raggedleft\arraybackslash}p{#1}}

\def\URuSi/{URu$_2$Si$_2$}
\def\q/{\textbf{q}}
\def\eps/{\bm{\epsilon}}
\def\qp/{\textbf{q}$||$}
\def\absq/{9.6\,\AA}
\def\°/{$^\circ$}
\def\O45/{U\,$O_{4,5}$}
\def\i/{\imath}
\def\a/{$\hat{\text{a}}$}
\def\b/{$\hat{\text{b}}$}
\def\c/{$\hat{\text{c}}$}
\def\SO/{spin-orbit}
\def\D4h/{D$_\text{4h}$}

\begin{document}
\title{Direct bulk sensitive probe of 5$f$ symmetry in \URuSi/}

\author{Martin~Sundermann}
  \affiliation{Institute of Physics II, University of Cologne, Z{\"u}lpicher Stra{\ss}e 77, 50937 Cologne, Germany}
\author{Maurits~W.~Haverkort}
  \affiliation{Max Planck Institute for Chemical Physics of Solids, N{\"o}thnizer Stra{\ss}e 40, 01187 Dresden, German}
\author{Stefano~Agrestini}
  \affiliation{Max Planck Institute for Chemical Physics of Solids, N{\"o}thnizer Stra{\ss}e 40, 01187 Dresden, German}
\author{Ali~Al-Zein}
	\altaffiliation[Present address: ]{Physics Department, Faculty of Science, Beirut Arab University (BAU), Beirut, Lebanon}
  \affiliation{European Synchrotron Radiation Facility (ESRF), B.P. 220, 38043 Grenoble C\'edex, France}
\author{Marco~Moretti~Sala}
  \affiliation{European Synchrotron Radiation Facility (ESRF), B.P. 220, 38043 Grenoble C\'edex, France}
\author{Yingkai~Huang}
  \affiliation{Van der Waals-Zeeman Institute, University of Amsterdam, Science Park 904, 1098 XH Amsterdam, Netherlands}
\author{Mark~Golden}
  \affiliation{Van der Waals-Zeeman Institute, University of Amsterdam, Science Park 904, 1098 XH Amsterdam, Netherlands}
\author{Anne~de~Visser}
  \affiliation{Van der Waals-Zeeman Institute, University of Amsterdam, Science Park 904, 1098 XH Amsterdam, Netherlands}
\author{Peter Thalmeier}
  \affiliation{Max Planck Institute for Chemical Physics of Solids, N{\"o}thnizer Stra{\ss}e 40, 01187 Dresden, Germany}
\author{Liu Hao~Tjeng}
  \affiliation{Max Planck Institute for Chemical Physics of Solids, N{\"o}thnizer Stra{\ss}e 40, 01187 Dresden, Germany}
\author{Andrea~Severing}
  \affiliation{Institute of Physics II, University of Cologne, Z{\"u}lpicher Stra{\ss}e 77, 50937 Cologne, Germany}

\begin{abstract}
The second-order phase transition into a hidden order phase in URu$_2$Si$_2$ goes along with an order parameter which is still a mystery, despite 30 years of research. However, it is understood that the symmetry of the order parameter must be related to the symmetry of the low lying local electronic $f$-states. Here we present results of a novel spectroscopy, namely core-level non-resonant inelastic x-ray scattering (NIXS). This method allows for the measurement of local high-multipole excitations and it is bulk sensitive. The observed anisotropy of the scattering function unambiguously shows that the 5$f$ ground state wave function is composed mainly, but essentially not purely, of the $\Gamma_1$ with majority $J_z$\,=$|4\rangle$\,+\,$|-4\rangle$ and/or $\Gamma_2$ singlet states. 
\end{abstract}

\maketitle

\section{Introduction}
The hidden order problem in URu$_2$Si$_2$ is an unanswered question in the field of strongly correlated electron materials. Although it is studied since several decades there is still no consensus about how this new phase forms. Understanding the hidden order phase formation is not only an intellectual problem, it will also advance concepts for designing quantum materials with new exotic properties. Many hidden order scenarios are based on the assumption of certain ground state symmetries and the present study addresses this aspect. A novel spectroscopic technique, non-resonant inelastic x-ray scattering (NIXS), that has become available thanks to high brilliance synchrotrons, allows to measure directly in a bulk sensitive experiment the symmetry of the 5$f$ ground state wave function in URu$_2$Si$_2$.

In heavy fermion rare earth or actinide compounds, the $f$ electrons are well localized at high temperatures, but as temperature is lowered hybridization with conduction electrons becomes increasingly effective, resulting in a more itinerant $f$-electron character at low temperatures. These hybridized $f$ electrons form narrow bands and have large effective masses. Quasiparticle interaction effects in these narrow bands are responsible for the many exciting phenomena present in heavy fermion compounds: multipolar order\,\cite{Santini2009}, unconventional superconductivity\,\cite{Pfleiderer2009} or quantum criticality\,\cite{Hilbert2007}. The hidden order phase in \URuSi/ is one example of the exotic low temperature phases found in this material class. \URuSi/ is a tetragonal heavy fermion compound that undergoes two phase transitions, the nonmagnetic \textsl{hidden order} ($HO$) transition at $T_{HO}$\,=\,17.5\,K that goes along with an appreciable loss of entropy, and a superconducting one at about 1.5\,K\,\cite{Palstra1985,Schlabitz1986,Maple1986,Kasahara2007}. Below the $HO$ transition small ordered magnetic moments were observed in the earlier studies, but turned out later to belong to a parasitic minority phase. With applied pressure (p\,$\ge$\,0.7\,GPa) the $HO$ order is replaced by an antiferromagnetic phase with large ordered moments (so-called $LMAF$-phase)\,\cite{Amitsuka2007}. The order parameter of the $HO$ phase has been subject of intense investigations since more than 30 years, but so far it remained hidden which has been the inspiration for its name. This second order transition into an electronically ordered state involves a reconstruction of the Fermi surface\,\cite{Meng2013,Bareille2014} and a change of quasiparticle scattering rate\,\cite{Chatterjee2013}. The Fermi surfaces of the $HO$ and high pressure $LMAF$ phase are very similar\,\cite{Hassinger2010}.

In \URuSi/ three energy scales have been identified: a hybridization gap of $\Delta_{hyb}$\,$\approx$\,13\,meV [150\,K]\,\cite{Park2012} that opens below 27\,K, another gap that opens in the $HO$ phase with $\Delta_{HO}$\,$\approx$\,4.1\,meV [50\,K] in the charge\,\cite{Aynajian2010,Schmidt2010,Meng2013,Bareille2014} as well as spin channel\,\cite{Broholm1991,Wiebe2007} and a resonance mode that appears in the $HO$ gap at $\cong$1.6\,meV [18\,K], also in both channels\,\cite{Buhot2014,Kung2015,Bourdarot2010}. Furthermore with entering the $HO$ phase the breaking of the fourfold rotational symmetry has been reported from torque experiments\,\cite{Okazaki2011} and high resolution x-ray diffraction on high quality crystals\,\cite{Tonegawa2007}. For a more detailed experimental and theoretical survey of physical properties of \URuSi/ we refer to the review article by Mydosh and Oppeneer\,\cite{Mydosh2011}. 

In intermetallic actinide compounds the valence state is often intermediate, and indeed, $N$-edge sum rules\,\cite{Jeffries2010}, life time reduced L-edge absorption\,\cite{Booth2016} and soft photoelectron spectroscopy measurements\,\cite{Fujimori2016} find a valence between 3$^+$ and 4$^+$ for \URuSi/. It is an itinerant system, and yet, electron correlations on the U atom will reduce the charge fluctuations and favor also a particular local irreducible representation\,\cite{Zwicknagl2003}. In this respect it is suggestive to assume that the U$^{4+}$\,$f^2$ configuration will give the dominant contribution which is in line with first-principle DMFT calculations\,\cite{Haule2009}. The question is now which of the U$^{4+}$\,($f^2$) states build up the itinerant state and lead to the formation of the $HO$. The present work presents the asymmetry of the inelastic x-ray scattering function S($\vec{q}$,$\omega$) as measured in a bulk sensitive, non-resonant inelastic x-ray scattering experiment (NIXS) and gives direct and quantitative information on the 5$f$ symmetry in URu$_2$Si$_2$. 

\begin{figure}
	\centerline{\includegraphics[width=0.96\columnwidth]{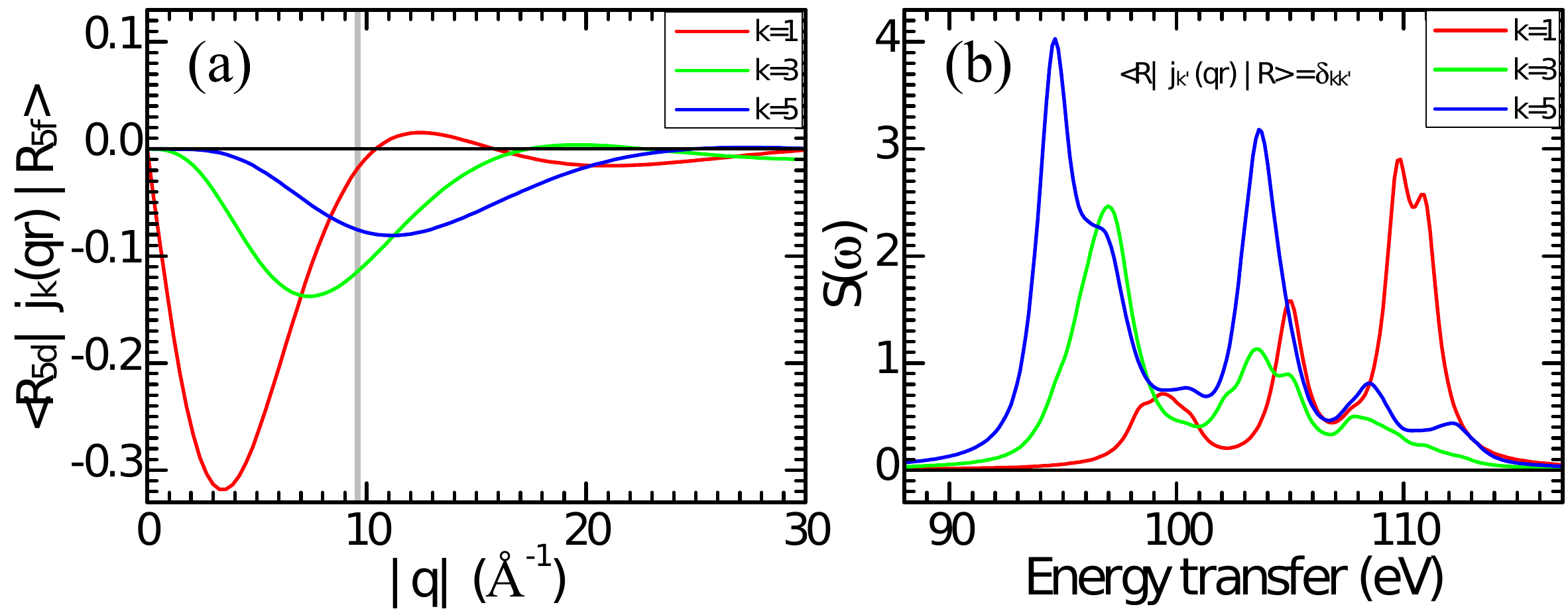}}
	\caption{Momentum $|\vec{q}|$ dependence (left) and energy dependence (right) of the scattering function S($\vec{q}$,$\omega$) at the U $O_{4,5}$-edge for dipole (k=1), octupole (k=3), and dotriacontapole (k=5) scattering orders. The gray vertical line marks the $\left|\vec{q}\right|$-range of the experiment. Note: features above $\approx$106\,eV appear unrealistically narrow since the proximity of continuum states is not accounted for.}
	\label{Fig1}
\end{figure}

To address the local 5$f$ degrees of freedom of URu$_2$Si$_2$ we will make use of the crystal-electric field (CEF) description of the U$^{4+}$\,$f^2$ configuration in $D_{4h}$ symmetry. The CEF splits the nine fold degenerate $J$\,=4 multiplet into five singlets and two doublets which can be written in the following way when using the $J_z$ representation. 
\begin{eqnarray*}
\Gamma_1^{(1)}(\theta) = \cos(\theta) \, | 0 \rangle + \sin(\theta) \sqrt{\frac{1}{2}} ( | 4 \rangle + | -4 \rangle )  \\
\Gamma_1^{(2)}(\theta) = \sin(\theta) \, | 0 \rangle - \cos(\theta) \sqrt{\frac{1}{2}} ( | 4 \rangle + | -4 \rangle )  \\
\Gamma_2 = \sqrt{\frac{1}{2}} ( | 4 \rangle - | -4 \rangle ) \\
\Gamma_3 = \sqrt{\frac{1}{2}} ( | 2 \rangle + | -2 \rangle ) \\
\Gamma_4 = \sqrt{\frac{1}{2}} ( | 2 \rangle - | -2 \rangle ) \\
\Gamma_5^{(1)}(\phi) = \cos(\phi) \, | \mp 1 \rangle + \sin(\phi) \, | \pm 3 \rangle \\
\Gamma_5^{(2)}(\phi) = \sin(\phi) \, | \mp 1 \rangle - \cos(\phi) \, | \pm 3 \rangle  
\end{eqnarray*}
Here the values $\theta$ and $\phi$ define the mixing of states that have equal irreducible representation, that is the singlet states $\Gamma_1^{(1,2)}$ and doublet states $\Gamma_5^{(1,2)}$. The phase relations between the $J_z$ states are defined such that the operator $\hat{J_x}$ is non-negative. Note, that $\Gamma_1^{(1)}$(90\°/)\,=\,-$\Gamma_1^{(2)}$(0\°/) and $\Gamma_5^{(2)}$(90\°/)\,=\,$\Gamma_5^{(1)}$(0\°/) and, depending on the mixing angles $\phi$ and $\theta$, the CEF states correspond to pure $J_z$ states 
($\Gamma_1^{(1)}$(90\°/)\,$\Leftrightarrow$\,$|4\rangle$+$|-4\rangle$, 
$\Gamma_2$\,$\Leftrightarrow$\,$|4\rangle$-$|-4\rangle$, 
$\Gamma_1^{(2)}$(90\°/)\,$\Leftrightarrow$\,$|0\rangle$, 
$\Gamma_5^{(1)}$(90\°/)\,$\Leftrightarrow$\,$|\pm 3\rangle$, and 
$\Gamma_5^{(2)}$(90\°/)\,$\Leftrightarrow$\,$|\pm 1\rangle$). 

Determining CEF excitations and their symmetry in intermetallic U compounds is by no means trivial since the 5$f$ electrons are more itinerant than e.g. the 4$f$ electrons in the rare earth series, and the classical tool --\,inelastic neutron scattering\,-- fails to observe sharp CEF excitations\,\cite{Broholm1987} due to dispersive effects and the large intrinsic widths that goes along with itinerant states. Nevertheless there have been many experimental and also theoretical attempts to determine the symmetries of the 5$f$ ground state and low lying electronic states in \URuSi/, and in literature a wide spectrum of different scenarios can be found. The anisotropy of the static susceptibility is well described with a $\Gamma_1^{(1)}$ singlet ground state, a $\Gamma_2$ as a first excited state and the next states above 15\,meV [170\,K]\,\cite{Nieuwenhuys1987}. Analyses of elastic constant measurements find similar results\,\cite{Yanagisawa2013a,Yanagisawa2013}. Also Kiss and Fazekas\cite{Kiss2005}, Hanzawa\,\cite{Hanzawa2012} and Kusunose \textsl{et al.}\,\cite{Kusunose2011} favour a $\Gamma_1^{(1)}$, the model of Kiss and Fazekas being also compatible with a $\Gamma_1^{(2)}$ singlet ground state\,\cite{Kiss2005}, but they all propose different first excited states from their theoretical considerations. Haule and Kotliar\,\cite{Haule2009} also propose two low lying singlet states, a $\Gamma_2$ singlet ground state and a $\Gamma_1^{(2)}$ as first excited state, a scenario that is compatible with the interpretation of polarized Raman studies that find a resonance at 1.6\,meV in the $A_{2g}$ channel in the $HO$ phase\,\cite{Buhot2014,Kung2015} \cite{footnote1}. Thermodynamic measurements by Santini and Amoretti\,\cite{Santini1994} and resonant x-ray scattering data by Nagao and Igarashi\,\cite{Nagao2005} are interpreted in terms of a $\Gamma_3$-singlet ground state with the $\Gamma_1^{(1)}$ as first excited state or alternatively with a $\Gamma_5^{(1)}$ ground state\,\cite{Nagao2005}. Another elastic constant study by Kuwahara \textsl{et al.}\,\cite{Kuwahara1997} yields a $\Gamma_4$ as lowest state. $\Gamma_5^{(1)}$ and $\Gamma_5^{(2)}$ doublets as ground states are concluded by thermodynamic studies of diluted \URuSi/\,\cite{Amitsuka1994} and theoretical considerations by Ohkawa and Shimizu\,\cite{Ohkawa1999} and Chandra et al.\,\cite{Chandra2013}. Finally $O$-edge x-ray absorption measurements by Wray \textsl{et al.}\,\cite{Wray2015} favour the $\Gamma_5^{(1)}$ and Sugiyama \textsl{et al.} the $\Gamma_5^{(2)}$ doublet\,\cite{Sugiyama1999} as ground state. 

\begin{figure}[h!]
	\centerline{\includegraphics[width=1.0\columnwidth]{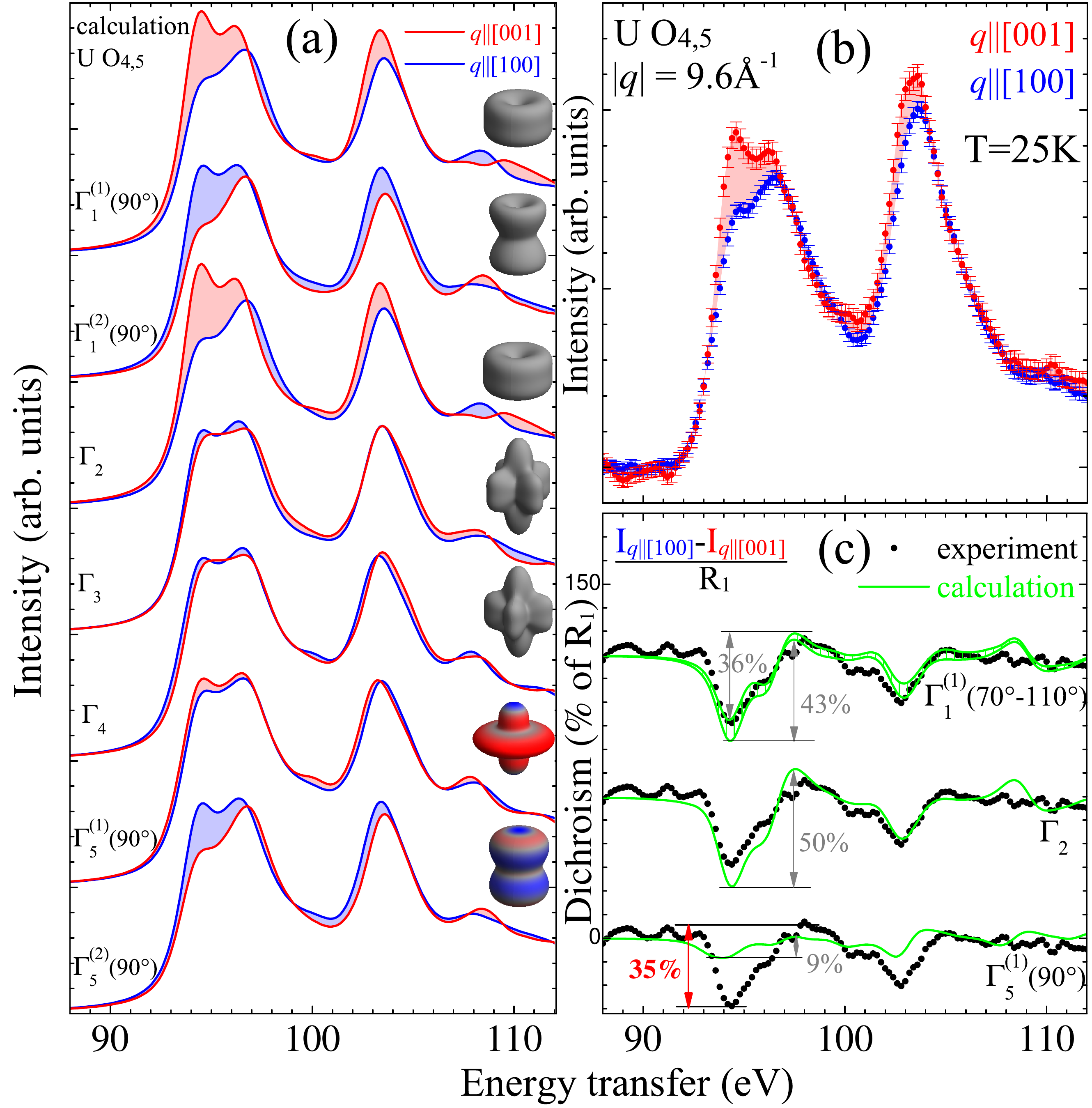}}
	\caption{(a)\,-\,(c): NIXS measurements of the U $O_{4,5}$-edge for $\left|\textbf{q}\right|$\,=\,9.6\AA$^{-1}$ and corresponding calculations for 5$d^{10}$4$f^2$\,$\rightarrow$\,5$d^{9}$4$f^3$. 
(a) Simulation of S($\vec{q}$,$\omega$) of U crystal-field states for $J$\,=\,4 in $D_{4h}$ symmetry for the two directions $\hat{q}$$\|$[100] (blue) and [001] (red). The insets show the corresponding electron densities (see section \textsl{Charge Density Plots} in the Appendix). 
(b) NIXS data for momentum transfers $\hat{q}$$\|$[100] (blue) and [001] (red) at T\,=\,25\,K. 
(c) Dichroism at 25\,K in \% defined as difference I$_{\hat{q}}$$_{\|[100]}$-I$_{\hat{q}}$$_{\|[001]}$ relative to peakhight R$_1$ as defined in the isotropic data (see Fig.\,3), data (black dots) and calculations (green lines) for the crystal-field states with the correct sign of dichroism. Here the data points have been convoluted with a Gaussian of 0.5\,eV FWHM.}
	\label{Fig2}
\end{figure}

\begin{figure}
  \centerline{\includegraphics[width=0.6\columnwidth]{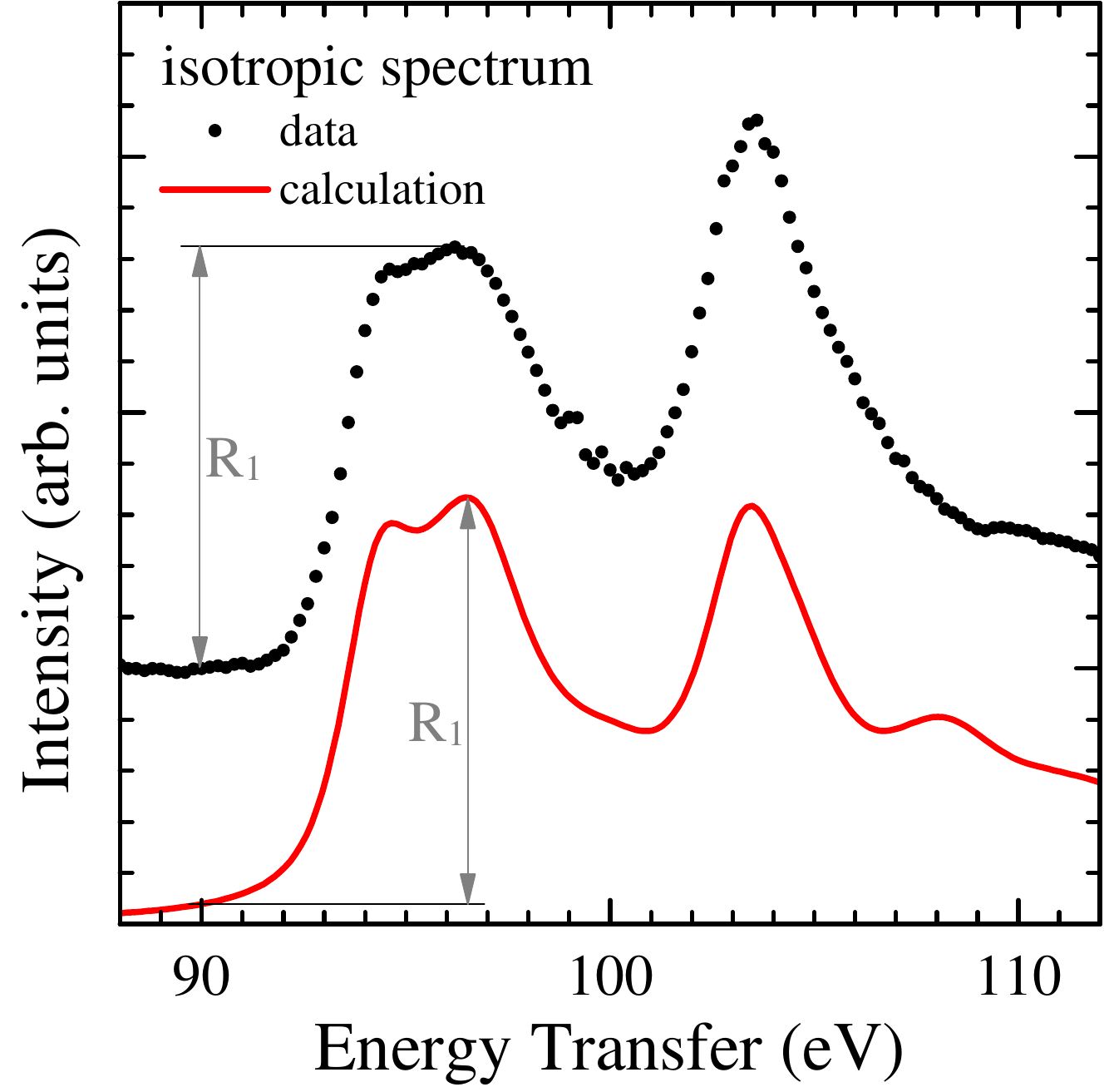}}	
  \caption{Experimental (black dots) and simulated (red line) isotropic spectrum of URu$_2$Si$_2$ at
the U $O_{4,5}$ edge (5$d^{10}$4$f^2$\,$\rightarrow$\,5$d^{9}$4$f^3$) for T\,$\le$\,25K. For details see text.}
	\label{Fig4}
\end{figure}

There is clearly room for clarification. Hence we aim at determining the symmetries of the ground state and low lying states in \URuSi/ using a spectroscopic method that directly probes the U 5$f$ shell. We performed a core-level \textsl{non-resonant} inelastic x-ray scattering experiment (NIXS) at the U $O_{4,5}$ edges ($5d$\,$\rightarrow$\,$5f$) with hard x-rays ($\approx$10\,keV) and large momentum transfers ($\left|\textbf{q}\right|$\,$\approx$\,9.6\,\AA$^{-1}$). NIXS is a photon-in-photon-out technique that was used in the recent past on single crystals for determining the wave functions of cerium based systems\,\cite{Willers2012,Rueff2015,Sundermann2015}. In NIXS the direction dependence of the momentum transfer \textbf{$\vec{q}$} is used in analogy to the linear polarization dependence in an x-ray absorption spectroscopy (XAS) experiment (see e.g.\ Ref.\,\cite{Hansmann2008} and also \cite{Wray2015}) and accordingly \textsl{multipole selection rules} give access to the ground state symmetry (dipole for XAS). The higher multipoles that contribute significantly to the scattering function S($\vec{q}$,$\omega$) at large momentum transfers contain more information than dipole so that e.g. asymmetries  with higher than twofold rotational symmetry can be detected\,\cite{Gordon2009,Willers2012}. In addition, at the U $O_{4,5}$-edge these excitations are significantly narrower than the dipole signal which is strongly broadened due to the proximity of continuum states\,\cite{SenGupta2011}. Most importantly, it should be mentioned that a NIXS experiment does not involve an intermediate state so that the quantitative modeling is as straightforward as for XAS and the use of hard x-rays makes the signal truly bulk sensitive in contrast to a soft XAS or soft RIXS experiment. 

\begin{figure}[h]
\begin{center}
\includegraphics[width=0.96\columnwidth]{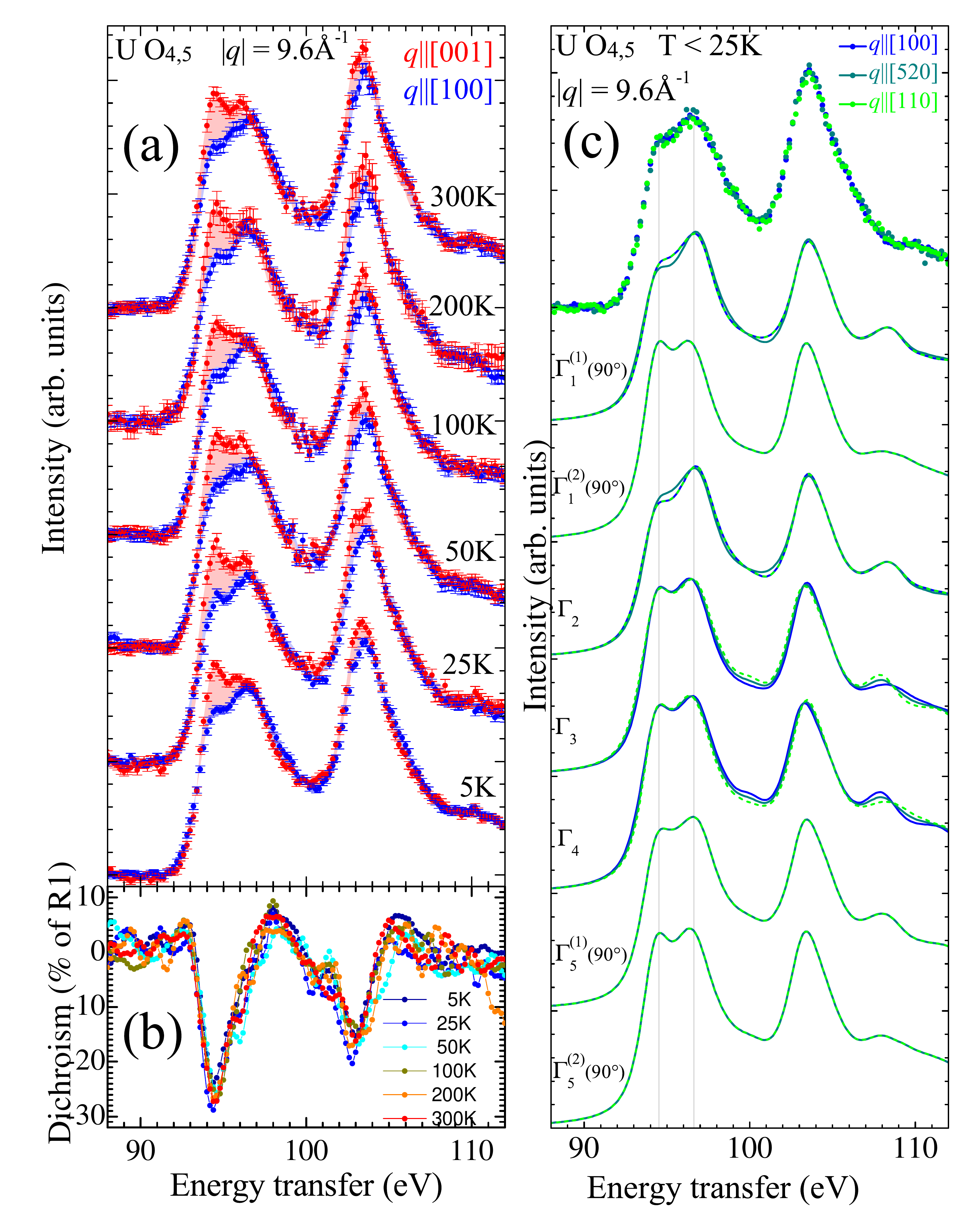}
\end{center}
	\caption{(a) Temperature dependence of the \URuSi/ \O45/ edge NIXS spectra for $\hat{q}$$\|$[100] (blue dots) and $\hat{q}$$\|$[001] (red dots). For better comparison, the $T$ dependent data are underlaid with the difference in spectral weight of the 5\,K data. (b) Dichroism I$_{\hat{q}}$$_{\|[100]}$-I$_{\hat{q}}$$_{\|[001]}$ for all temperatures, convoluted with a Gaussian of 0.5\,eV FWHM. (c) NIXS data and simulations (5$d^{10}$4$f^2$\,$\rightarrow$\,5$d^{9}$4$f^3$) for in-plane momenta $\hat{q}$ parallel to [100] and for $\hat{q}$ turned towards [010] by 22.5\°/ and 45\°/, i.e. $\hat{q}$$\|$[100] (blue), $\hat{q}$$\|$[520] (dark green) and $\hat{q}$$\|$[110] (light green) for all CEF states. }
	\label{Fig3}
\end{figure}

\section{Results}
Several NIXS studies, also on uranium compounds, show experimentally and theoretically how the multiplet excitations develop with increasing momentum transfer\,\cite{Soininen2005,Gordon2008,Rueff2010,Bradley2010,Caciuffo2010,SenGupta2011,Willers2012,Laan2012,Huotari2015}. However, for convenience of the reader we recapitulate briefly the principle of NIXS: when working at large momentum transfers the expansion of the transition operator exp(i$\vec{q}$$\vec{r}$) in spherical harmonics cannot be truncated after the first term, thus giving rise to excitations due to quadru-, octupole and higher order contribution in S(\textbf{$\vec{q}$},$\omega$). Figure\,\ref{Fig1} shows the three non vanishing contributions to S($\vec{q}$,$\omega$) calculated for the U $O_{4,5}$-edge; the radial part as function of momentum transfer in (a) and the isotropic spectra in (b), each for the dipole and higher multipole contributions. The excitations due to scattering from higher multipoles contribute substantially to the total intensity already for momentum transfers of $\left|\textbf{q}\right|$\,$\approx$\,9.6\,\AA$^{-1}$. By performing such an experiment on a single crystal and measuring the \textbf{q}-direction dependence will give S($\vec{q}$,$\omega$). This then can be used for the CEF analysis where each state will have a specific direction dependence.

Figure\,\ref{Fig2}(a) shows the simulation of S($\vec{q}$,$\omega$) of the $5d$\,$\rightarrow$\,$5f$ transition ($O_{4,5}$-edge) for the nine states of the $J$\,=\,4 ground state multiplet. Contributions from other valence configurations are neglected. For more detailed information about the simulation we refer to the section \textsl{Methods}. The spectra corresponding to the respective CEF states are calculated for the two directions \textbf{$\hat{q}$}$\|$[100] and \textbf{$\hat{q}$}$\|$[001] and some of them exhibit a strong direction dependence. Here $\theta$ and $\phi$ are chosen such that the anisotropies are maximum, i.e. for the extreme cases of pure $J_z$ states (see definition of CEF states). The insets in Fig.\,\ref{Fig2}(a) show the respective two electron 5$f$ charge densities. The charge densities of the pure states in Fig.\,\ref{Fig2}(a) that appear rotational invariant do show lobes for $\theta$ and $\phi$\,$\neq$\,0 or 90\°/ (see Fig.\,\ref{FigS1} and Fig.\,\ref{FigS2} in Appendix).

The NIXS experiment was performed at large momentum transfers (see Experimental set-up in section \textsl{Methods}) so that the signal is dominated by higher order scattering (\textsl{beyond dipole}). Data were taken below and above the $HO$ transition at 5\,K and 25\,K and with successively rising temperature up to 300\,K. All data shown are background corrected. 

Before discussing the direction dependence, we first show in Fig.\,\ref{Fig4} the isotropic data (see section \textsl{Isotropic spectra} in the Appendix for construction of isotropic spectra) together with a simulation using the ionic 5$f^2$ configuration for the U. We can clearly observe a very good agreement, thereby establishing that the spectrum is dominated by the atomic multiplet structure. This is important since this validates a posteriori the use of local probes (such as core level NIXS) to test models utilizing also local irreducible representations. Perhaps more surprising is that a single ionic configuration can reproduce the spectrum so well despite the known covalency of the U. However, it has been explained by, for example, Gunnarrson and Sch\"onhammer \cite{Gunnarsson1983} as well as deGroot \cite{deGroot1994}, that $d\,\rightarrow\,f$ and $p\,\rightarrow\,d$ core-level XAS (NIXS) for 4$f$/5$f$ and 3$d$ compounds, respectively, highlights the spectral weight of the energetically lowest lying (and major) configuration at the expense of those of the higher lying configurations, making the technique extremely powerful for determining the symmetry of the ground state (see \textsl{Spectroscopy} in the Appendix).

We now discuss the direction dependence of the data measured above the $HO$ transition since they are not affected by any possible impact of the $HO$. In Fig.\,\ref{Fig2}(b) the NIXS data of \URuSi/ at 25\,K are shown for the in-plane direction \textbf{$\hat{q}$}$\|$[100] (blue dots) and out-of-plane direction \textbf{$\hat{q}$}$\|$[001] (red dots). The error bars reflect the statistical error. There is a large anisotropy that can be directly compared with our simulations. 

A more detailed comparison of data and simulations excludes immediately the $\Gamma_1^{(1,2)}$($\theta$) states with strong $J_z$\,=\,$|0\rangle$ contributions, the $\Gamma_3$ and $\Gamma_4$ singlets with $J_z$\,=\,$|2\rangle$ and $J_z$\,=\,$|-2\rangle$ as well as the $\Gamma_5^{(1,2)}$($\phi$) states with strong $J_z$\,=\,$|\pm1\rangle$ weight. Actually, only singlet states with majority $|+4\rangle$ and $|-4\rangle$ or a doublet with majority $|\pm3\rangle$ show the correct direction dependence, i.e. red over blue (see Fig.\,\ref{Fig2}(a)). To be more quantitative we compare the measured dichroism of about 35\% (see Fig.\,\ref{Fig2}(c)) with the simulated dichroism of the \URuSi/ wave functions in question. Here the dichroism is defined as the difference of the intensities for {\textbf{$\hat{q}$}$\|$[100]} and {\textbf{$\hat{q}$}$\|$[001]} relative to the peak height R$_1$ with R$_1$ being the intensity difference of pre-edge at $\approx$90\,eV and peak height at $\approx$96\,eV of the isotropic spectrum (see Fig.\,4, described in section \textsl{Appendix}). We find that $\Gamma_1^{(1)}$(90\°/) (or $\Gamma_1^{(2)}$(0\°/)) and also $\Gamma_2$ reproduce the size of the anisotropy quite well, although their dichroism is with 43\% or 50\% slightly larger than the measured value. A $\Gamma_1^{(1)}$ state of majority $J_z$\,=\,$|4\rangle$ and $|-4\rangle$ symmetry, but with some $J_z$\,=\,$|0\rangle$ ($\Gamma_1^{(1)}$(70\°/ or 110\°/)) would produce a slightly smaller dichroism of about 36\% (see Fig.\,2(c)). 

The $\Gamma_5^{(1)}$(90\°/) (or $\Gamma_5^{(2)}$(0\°/)), i.e. the doublet states with the highest amount of $J_z$\,=\,$|\pm3\rangle$, do not yield sufficient dichroism: the dichroism of 9\% is by a factor four too small and would decrease further or even change sign with increasing amount of $|\pm1\rangle$. (In the following we skip writing out the $\Gamma_i^{(2)}$ alternative state because of $\Gamma_i^{(1)}$(90\°/)\,=\,-$\Gamma_i^{(2)}$(0\°/), i\,=\,1 or 5.)  

Figure\,\ref{Fig3}(a) shows the $ac$ asymmetry of the scattering function for all temperatures. Also here the error bars reflect the statistical error.  We find that within the error bars the 5\,K and 25\,K are identical. We further find that there is no change with temperature up to 300\,K as is demonstrated by plotting the dichroism for all temperatures in Fig.\,\ref{Fig3}(b). The Boltzmann population with temperature of any state other than the $\Gamma_1^{(1)}$(70\°/-90\°/-110\°/) and $\Gamma_2$ state will change the direction dependence of the scattering (compare Fig.\,\ref{Fig2}(a)). Hence we conclude from the absence of any changes in the spectra up to 300\,K that the ground state consists mainly of the $\Gamma_1^{(1)}$(70\°/-90\°/-110\°/) or the $\Gamma_2$ singlet, or that one of the two singlets forms the ground state with the respective other state close in energy. We can further estimate from the impact of thermal occupation that the states with weak dichroism like the $\Gamma_5^{(1)}$(90\°/), $\Gamma_3$ and $\Gamma_4$ must be higher than 150\,K (13\,meV) while states with stronger opposite anisotropy must be even higher in energy.

Figure\,\ref{Fig3}(c) shows data taken within the plane, for \textbf{$\hat{q}$}$\|$[100] and for two directions 22.5\°/ and 45\°/ towards [010] as well as the respective simulations for all CEF states. Neither below nor above the $HO$ order transition we can resolve any anisotropy within the statistical error bar. This is not in contradiction with our previous findings that either one of the two singlet states $\Gamma_1^{(1)}$(70\°/-90\°/-110\°/) and $\Gamma_2$ forms the ground state since the asymmetries expected from simulations are rather small and most likely covered by statistics of this low count experiment. The in-plane data even confirm the out-of-plane data when comparing the measured and simulated shape of the spectra: for example, the peak at 94\,eV is clearly smaller than the peak at 97\,eV for the simulated $\Gamma_1^{(1)}$(90\°/) and $\Gamma_2$ spectra (see gray lines in Fig.\,\ref{Fig3}(c)) in agreement with the data while the two peaks are about the same for all other states. For the in-plane simulation for different values of $\theta$ and $\phi$ we refer to Fig.\,\ref{Fig6} (see Appendix).

\section{Discussion}
Our results of a ground state that consists mainly of $\Gamma_1^{(1)}$(70\°/-90\°/-110\°/) and/or $\Gamma_2$ agree well with the description of the anisotropy of the static susceptibility\,\cite{Nieuwenhuys1987} and the analysis of the temperature dependence of the elastic constants\,\cite{Yanagisawa2013a,Yanagisawa2013} which are well described with a $\Gamma_1^{(1)}$ of majority $J_z$\,=\,+\,4 and -\,4 , a $\Gamma_2$ as first excited state and another state above 150\,K. It also confirms DMFT calculations that finds these two singlet states as low lying states close in energy (see Ref.\,\cite{Haule2009} and also in Supplementary of Kung \textsl{et al.}\,\cite{Kung2015}) but the experiment yields the additional information that the $J_z$\,=\,+\,4 and -\,4 in the $\Gamma_1$ is dominating. We further would like to stress that linear polarized XAS data at the U $O_{4,5}$ edge\,\cite{Wray2015} also agree with our findings in the sense that both, the NIXS \textsl{and} XAS dichroism, rule out the  $\Gamma_1^{(2)}$(90\°/) (or  $\Gamma_1^{(1)}$(0\°/)), the $\Gamma_3$ or the $\Gamma_4$ as possible ground states and find no temperature dependence across the $HO$ transition. The smaller direction dependence that lets the authors of Ref.\,\cite{Wray2015} assign the $\Gamma_5^{(1)}$ doublet as ground state might be due to the higher surface sensitivity of the XAS experiment. 

A pure $\Gamma_1^{(1)}$(70\°/-90\°/-110\°/) or $\Gamma_2$ or both close in energy does confront us with the dilemma that neither would break the $C_4$ in-plane symmetry as suggested by the torque\,\cite{Okazaki2011}, high precision x-ray\,\cite{Tonegawa2007}, and elastoresistance\,\cite{Riggs2015} results nor would an ordering out of a singlet state yield sufficient loss of entropy across the $HO$ transition. To allow for a rank-5 $E^-$ $HO$ parameter as in the fully microscopic itinerant approach\,\cite{Ikeda2012,Thalmeier2013} the inclusion of the twofold degenerate CEF state of $E$-type is a necessity in the present more localized picture. Also the interpretation of the resonance intensity in the main A$_{2g}$ and other Raman channels in terms of a staggered chirality density wave requires a mixing of $\Gamma_1$ and $\Gamma_2$ singlet states that support a hexadecapole-type hidden order parameter. The model of Kung \textsl{et al.} contains both dominant A$_{2g}$ and subdominant B$_{1g}$ symmetry parts.  The latter involves higher energy CEF states and couples to the lattice leading to a secondary orthorhombic distortion that leaves only twofold symmetry. This causes a leakage of resonance intensity into forbidden channels \,\cite{Kung2015}.

Our experiment shows that the CEF components are mainly of the $\Gamma_1^{(1)}$(90\°/) or  $\Gamma_2$ singlet type but we did not observe the theoretically maximum possible dichroism (see Fig.\,\ref{Fig2}(c)) so that the data allow the presence of some other symmetry. The mixing of an irreducible representation, other than $\Gamma_1^{(1)}$(70\°/-90\°/-110\°/) or $\Gamma_2$, into the ground state cannot rely on Boltzmann occupation since that would have been observable in the temperature dependence of the NIXS data. However, a Kondo-type mechanism where an $f^3$ Kramers doublet hybridizes with the crystal-electric field manifold of the energetically more favorable $f^2\epsilon^{+1}_k$ configuration is feasible for constructing a ground state with different $f^2$ CEF characters. Here $\epsilon^{+1}_k$ denotes an electron in the host conduction band. We recall that a hybridization gap of 13\,meV opens up below 27\,K\,\cite{Park2012} which is also seen in the DMFT calculation that includes these Kondo processes\,\cite{Haule2009,Kung2015}. The stabilization energy of this Kondoesque wave function should be of the same order as the hybridization gap and the contributing CEF configurations should also be within this energy range. 

\section{Summary}
The bulk sensitive, U $O_{4,5}$ non-resonant inelastic x-ray scattering data of URu$_2$Si$_2$ exhibit the atomic multiplet structure of the $f^2$ configuration. The huge out-of-plane anisotropy shows that the symmetry of the ground state consists mainly of the $\Gamma_1^{(1)}$(90\°/) or $\Gamma_2$ singlet states in the U$^{4+}$\,($f^2$) configuration and/or that these two states are close in energy. The data do not exhibit any temperature dependence, neither across the $HO$ phase transition nor in the temperature interval up to 300\,K, the latter setting constraints to the proximity of next higher excited states. Scenarios for constructing a ground state that is a superposition of different irreducible representations without relying on Boltzmann statistics are discussed.

\section{Appendix}
\subsection{Samples}A high-quality single crystal of URu$_2$Si$_2$ was grown with the traveling zone method in the two-mirror furnace in Amsterdam under high purity (6N) argon atmosphere. The crystal was checked and oriented with x-ray Laue diffraction for its single-crystalline nature. The oriented crystal was cut using the spark erosion method after which the relevant surfaces [(100), (110), and (001)] were polished. A bar-shaped piece of the single crystal was characterized by resistance measurements. 

\subsection{Experimental set-up} The scattering function S($\vec{q}$,$\omega$) was measured in a non-resonant inelastic x-ray scattering experiment (NIXS) at the beamline ID20 at ESRF. Two monochromators [Si(111) and Si (311)] set the incident energy to 9690 eV and the scattered intensity was analyzed by one column of three Si(660) crystal analyzers at an in-plane scattering angle of 2$\vartheta$\,=\,153\°/ and detected in a \textsl{Maxipix} 2D detector with an overall energy resolution of about 0.8\,eV. This setting corresponds to a momentum transfer of $\left|\textbf{q}\right|$\,=\,9.6\,\AA$^{-1}$. The crystals with (100), (110), and (001) surfaces allowed realizing \textbf{$\hat{q}$}$\|$[100],[110], and [001] in specular geometry and also other directions when going off specular. It turned out that specular geometry (same path for photon in as for photon out) is not necessary since $\hat{q}$$\|$[110] measured specular on the (110) crystal and 45\°/ off specular on the (100) crystal gave the same result. For cooling the samples were mounted in a He flow cryostat. The elastic line was measured before each setting to determine the zero energy transfer and exact instrumental resolution for each analyzer. The spectra of the \O45/ edges were then normalized to their pre-edge intensity. Scans over a wide energy range were taken in order to correct for the Compton scattering and some minor constant background. The Compton background was fitted to a Gaussian and then subtracted from the data. 

\subsection{Spectroscopy} Why does $O$-edge XAS or NIXS resemble the $f^2$ multiplet structure and why is $O$-edge XAS or NIXS sensitive to the symmetry? The reason for this is that the energy order of the local configurations (in a configuration interaction picture) of the ground state problem and of the NIXS/XAS core-hole final state problem is identical. Hence, the spectral weights of the other local configurations are strongly suppressed due to quantum mechanical interference effects. This is e.g. well explained by O. Gunnarsson and K. Sch\"onhammer \cite{Gunnarsson1983} Fig.\,7 for the case of Ce $M_{4,5}$ XAS and by de Groot \cite{deGroot1994} pages 549 and 550 for the case of 3$d$ transition metal $L_{2,3}$ XAS. A quote from the latter: \textit{XPS is sensitive to the charge transfer effects \,.\,.\,.\,.\,.\,.\,.\,. while XAS is sensitive to the symmetry of the ground state with its characteristic multiplet.}  It is important that the relevant intra- and inter-shell Coulomb interactions are of similar size in order to have the same order of configuration energies in the ground state problem as well as  core-hole final state problem: this is true for the Ce $M_{4,5}$ (3$d$, 4$f$) , U $O_{4,5}$ (5$d$, 5$f$), transition metal $L_{2,3}$ (2$p$,3$d$) edges, but not for e.g. the Ce $L_{2,3}$ (2$p$,5$d$ with 4$f$ as spectator) or U $L_{2,3}$ (2$p$,6$d$ with 5$f$ as spectator) since here the Coulomb interaction of 5$d$-4$f$ or 6$d$-5$f$ is negligible in comparison with the 2$p$-4$f$ and 4$f$-4$f$ or 2$p$-5$f$ and 4$f$-5$f$, respectively. For further reading we refer to Ref.\,\cite{Gunnarsson1983} and \cite{deGroot1994}.

\subsection{Simulations} The simulations include \SO/ coupling and Coulomb interaction and are based on an ionic model with a U$^{4+}$\,5$f^2$ configuration. The atomic values are calculated with the Cowan code\,\cite{Cowan1981} but the Slater integrals for Coulomb interactions are reduced by a constant factor to account for the screening of the moments in the solid. The 5$f$--5$f$ and 5$d$--5$f$ reduction was adjusted to about 50\% to match the experimental energy spread of the multiplet signal of the isotropic data in Fig.\,S1 (for construction of isotropic spectrum see below). The ratio of multipole contributions was slightly adjusted by varying $|\q/|$\,\cite{Gordon2008}. In the simulations the actual value for $|\q/|$ was slightly larger than according to the experimental scattering triangle because the radial part of the wave functions that enter the calculations are based on the atomic values. For all finite values of \SO/ coupling and Coulomb interaction the $J$\,=\,4 multiplet forms the Hund's rule ground state. The relative contributions of different angular momenta $L$\,=\,3,4,5 depend on the ratio of \SO/ coupling and Coulomb interaction and are 1\%, 14\%, and 85\% for our reduction factors, respectively. 

Within the $J$\,=\,4 basis we create the local eigenstates ($\Gamma_1$ to $\Gamma_5$) restricted to the $f^2$ configuration by combining different states of $J_z$ considering the constraints by group theory. The Hamiltonian includes the local spin-orbit coupling and multipolar Coulomb interaction, which are much larger (up to 1 Rydberg) than the final-state core-hole lifetime (order of 1\,eV), but neglects the effects of crystal-field, covalent bonding and band-formation of the crystal, which will be smaller or of the same order of magnitude as the core-hole lifetime. The calculations are performed using the \textsl{Quanty} code\,\cite{Haverkort2012,Haverkort2016}.

To account for instrumental resolution, lifetime effects, and interference effects with the continuum the multiplet lines are broadened with a Gaussian (FWHM\,=\,0.8\,eV), a Lorentzian (FWHM\,=\,1.3\,eV) and a Mahan-type line shape (with an asymmetry factor 0.18 and an energy width of the continuum of 1000\,eV) in order to mimic the asymmetry due to the itinerancy. 

\subsection{Charge Densities} The charge densities are calculated for two electrons. The surface thus tells how much charge can be found in a specific direction. For a system with one electron this would be a plot of the orbital that is occupied. For a system that can be represented by a single Slater determinant it shows a "sum" of the different orbitals occupied. The color is related to the spin density with up as red, down as blue and zero as gray. 

For URu$_2$Si$_2$ often $LSJJ_z$ coupling is assumed whereby $L$ $S$ $J$ and $J_z$ are all good quantum numbers. Doing so results in density plots with much more features than in the present manuscript. This assumption basically is equivalent to saying that $F_2$, $F_4$, and $F_6$ Slater integrals are infinitely larger than the spin-orbit coupling. That approximation is not valid and spin-orbit coupling mixes states $LS$ with states $L$+1 $S$-1 or $L$-1 $S$+1. This mixing is included in our calculations. Furthermore we know that both the multipole part of the Coulomb interaction as well as the spin-orbit interaction are not (not really) screened in a solid. In other words the U 5$f$ shell in URu$_2$Si$_2$ is in-between $LS$ and $jj$ coupling and was taken in account.

\subsection{Isotropic spectrum:} The isotropic spectrum is given by the trace of the conductivity tensor. For dipole transitions (k\,=\,1) this tensor can be written as a 3x3 matrix with two independent diagonal elements in $D_{4h}$ symmetry. However, for higher multipoles the conductivity tensor has also a higher dimension, i.e.\ 7\,x\,7 for octupole (k\,=\,3) and 11\,x\,11 for dotriacontapole (k\,=\,5). Here we obtain the \textsl{experimental} isotropic spectrum containing these three relevant conductivity tensors by combining 10 independently measured directions. The \textsl{calculated} isotropic spectrum is obtained by averaging over all CEF states. The red line in Fig.\,\ref{Fig4} is the simulation of the isotropic data after optimizing the respective parameters. 

\subsection{$\theta$ and $\phi$ dependence of spectra}
{\raggedleft\textbf{Out-of-plane anisotropy:}} Figure\,S1(a) and (b) show the out-of-plane anisotropy in the U
$O_{4,5}$-edge NIXS spectra of the $\Gamma_1^{(1,2)}$($\theta$) and $\Gamma_5^{(1,2)}$($\phi$) wave functions for values of $\theta$ and $\phi$ between 0 and 90\°/. The insets show the respective charge densities. Please note that -$\Gamma_1^{(1)}$(90\°/)\,=\,$\Gamma_1^{(2)}$(0\°/) and $\Gamma_1^{(2)}$(90\°/)\,=\,$\Gamma_1^{(1)}$(0\°/). The same holds for the $\Gamma_5^{(1,2)}$, i.e. -$\Gamma_5^{(1)}$(90\°/)\,=\,$\Gamma_5^{(2)}$(0\°/) and $\Gamma_5^{(2)}$(90\°/)\,=\,$\Gamma_5^{(1)}$(0\°/).

\begin{figure}[h]
	\centerline{\includegraphics[width=0.98\columnwidth]{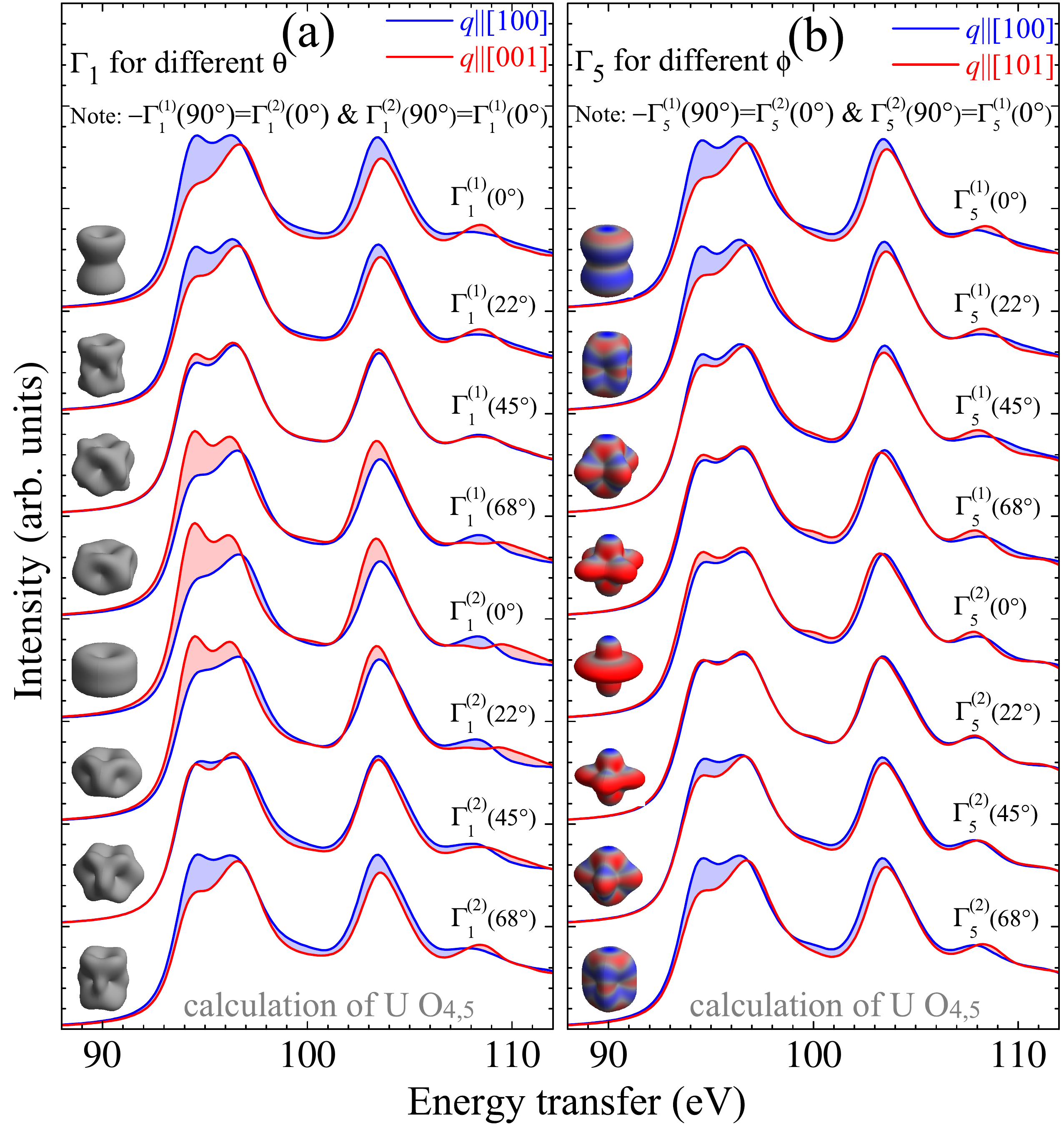}}
	\caption{NIXS simulations for momenta $\hat{q}$ parallel to [100] (blue) and [001]
(red) for the two states (a) $\Gamma_1^{(1,2)}$($\theta$) and $\Gamma_5^{(1,2)}$($\phi$)
for different $J_z$ admixtures expressed in terms $\theta$ and $\phi$}
	\label{FigS1}
\end{figure}

{\raggedleft\textbf{In-plane anisotropy:}} Figure\,S2(a) and (b) show the in-plane anisotropy in the U $O_{4,5}$-edge NIXS spectra of the $\Gamma_1^{(1,2)}$($\theta$) and $\Gamma_5^{(1,2)}$($\phi$) wave functions for values of $\theta$ and $\phi$ between 0 and 90\°/. The insets show the respective charge densities. According to simulations none of the states exhibits a noteworthy in-plane anisotropy in the scattering intensity although the charge densities appear anisotropic and although there are multipole contributions in S($\vec{q}$,$\omega$). This can be understood when considering that the scattering process at the U$_{4,5}$-edge is from 5$d$\,$\rightarrow$\,5$f$, i.e.\ we scatter from a state with a complicated shaped charge distribution to another one with a non-simple shape. 

\begin{figure}[h]
	\centerline{\includegraphics[width=0.98\columnwidth]{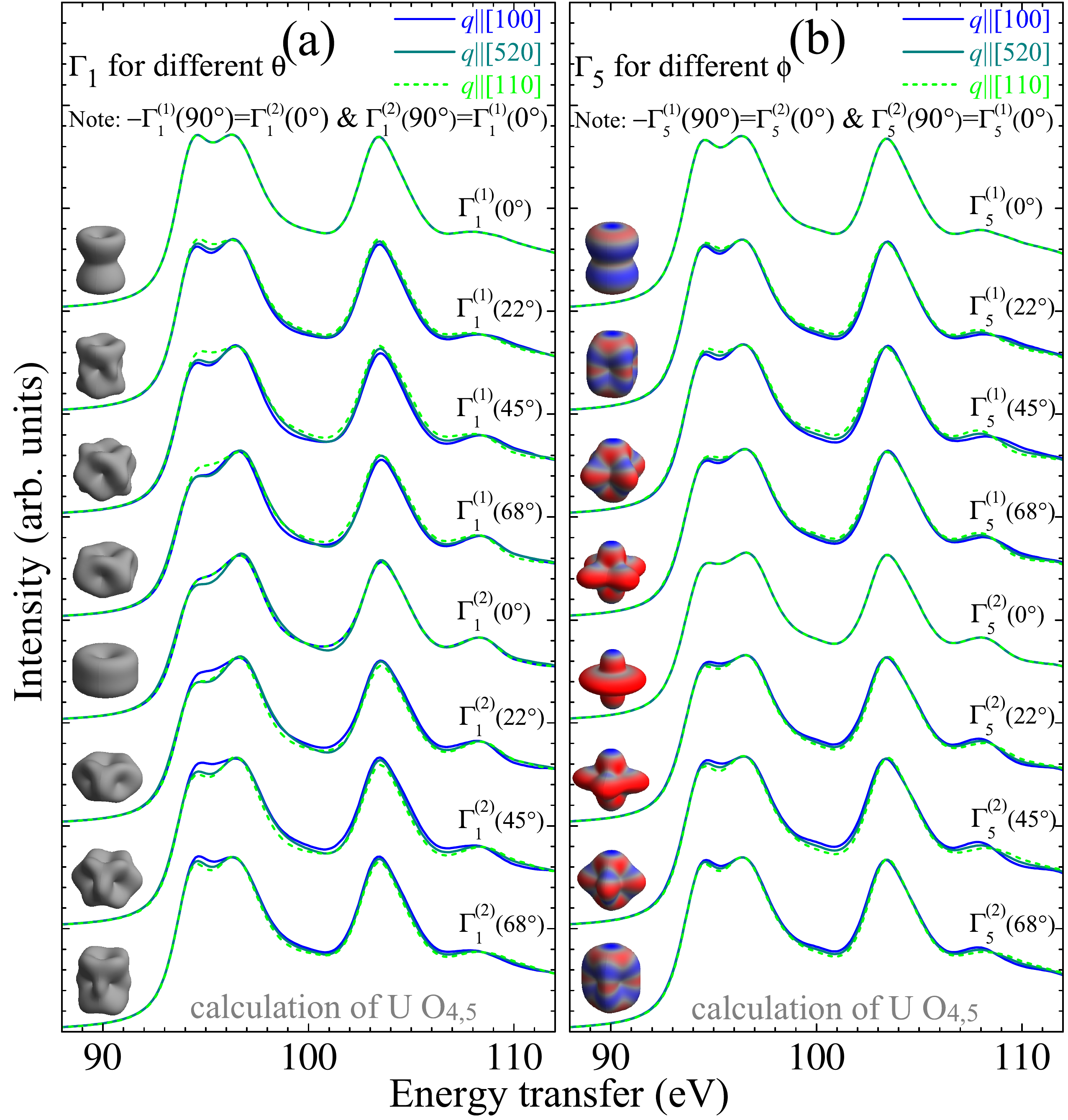}}
	\caption{NIXS simulations of U $O_{4,5}$ for momenta $\hat{q}$ parallel to [100] and for $\hat{q}$ turned towards [010] by 22.5\°/ and 45\°/, i.e. $\hat{q}$$\|$[100] (blue), $\hat{q}$$\|$[520] (dark green) and $\hat{q}$$\|$[110] (light green) for the two states (a) $\Gamma_1^{(1,2)}$($\theta$) and (b) $\Gamma_5^{(1,2)}$($\phi$) for different $J_z$ admixtures expressed in terms of $\theta$ and $\phi$.}
	\label{FigS2}
\end{figure}

\begin{acknowledgments}
We thank Areem Nikitin for for characterizing the sample by transport. M.S. and A.S. benefited from support of the German funding agency DFG (Project  600575). We further acknowledge ESRF for provision of synchrotron radiation facilities (proposal HC1533 and HC2252). 
\end{acknowledgments}

\bibliographystyle{apsrev4-1}

\end{document}